# Towards barrier free contact to MoS₂ using graphene electrodes

Yuan Liu[1,||], Hao Wu[1,||], Hung Chieh Cheng[1,||], Sen Yang[1], Enbo Zhu[1], Qiyuan He[2], Mengning Ding[2], Dehui Li[2], Jian Guo[1], Nathan O Weiss[1], Yu Huang[2,3], and Xiangfeng Duan[2,3,*]

[1]Department of Materials Science and Engineering, University of California, Los Angeles, CA 90095, USA; [2]Department of Chemistry and Biochemistry, University of California, Los Angeles, CA 90095, USA; [3]California Nanosystems Institute, University of California, Los Angeles, CA 90095, USA;

The two-dimensional (2D) layered semiconductors such as MoS₂ have attracted considerable interest for a new generation of ultrathin electronics and optoelectornics.[1-11] However, their promise has been largely stalled by the difficulties in making optimized metal contacts to these atomically thin materials. Considerable efforts have been devoted to mitigating the contact resistance by using low work function metal electrodes, high temperature annealing, or phase engineering.[12-20] Although these approaches have successfully reduced the contact resistance, it is not yet possible to achieve linear Ohmic contact particularly at low temperature, where a finite Schottky barrier dominates the carrier transport due to Fermi level pinning at conventional metal-MoS₂ interface[16]. This contact barrier remains a significant obstacle towards optimized device performance. Herein we present a new strategy by using graphene as a tunable contact for achieving Ohmic contact to 2D semiconductors. With a finite density of states, the Fermi level of graphene can be readily modified by gate potential to ensure a nearly perfect match with **MoS$_2$. Importantly, with a graphene contact, we demonstrate, for the first time, a transparent contact to MoS$_2$ with essentially zero contact barrier and linear output behaviour at cryogenic temperatures (down to 1.9 K). With the barrier-free carrier transparent contacts, we show that a metal-insulator-transition (MIT) can be observed in a two-terminal device, a phenomenon that could be easily masked by Schottky barrier and only seen in four-terminal devices in conventional metal-contacted MoS$_2$ system. Furthermore, benefited from the minimized contact barrier, our device display a record high extrinsic (two-terminal) field effect mobility over 1300 cm$^2$/ V s. We believe our strategy of using graphene as barrier free contact could shed light on contact engineering in atomically thin 2D transistors and other conventional transistors in general.**

To fabricate our device, two stripes of single layer graphene are first mechanical exfoliated on 300 nm thick silicon dioxide substrate and used as the back contact electrodes, while the channel length is defined by the distance between those two graphene strips (Figure 1a). Next, mechanical exfoliated MoS$_2$ stripe is directly transferred on top of two graphene electrodes, using a dry alignment transfer technique (see method). This direct integration of exfoliated graphene contact with exfoliated MoS$_2$ is essential here to avoid lithography/etching process that could introduce irremovable polymeric residues and adversely impact the charge transport across the graphene-MoS$_2$ vertical contact. The residue-free dry transfer method provide an atomically sharp and ultraclean interface between graphene and MoS$_2$ (as shown in TEM cross-section image in Figure 1b), which can ensure pure *van der Waals* contacts and minimize defects and charge trapping sites. Finally, Cr/Au (10/50 nm) electrodes is used to contact graphene with standard e-beam lithography, metal deposition and lift-off process, with the final device shown in Figure 1c.

Electrical transport studies are carried out under dark environment. Figure 2a and 2c shows standard I-V output characteristic at room temperature for monolayer and multilayer $MoS_2$ (20 to 30 layers) with a W/L ratio around 1, respectively. Without any post annealing process, our devices exhibit linear I-V output characteristic in both cases. The Ohmic behaviour remains stable with decreasing temperature down to 1.9 K (Figure 2b, d), suggesting a truly transparent and barrier free contact for $MoS_2$. This behaviour is observed in all our samples with additional data in supplementary information. **To the best of our knowledge, this low temperature Ohmic contact behaviour have never be achieved before.** The closest approach is applying Ti as contact metal with high-temperature post annealing[20], where obvious nonlinear behaviour is still observed at a temperature of 5 K.

The excellent contact in our devices may be attributed to two reasons. First, due to unique linear dispersion relationship with finite density of states in graphene, the Fermi level could be easily shifted by gate voltage or and nearby $MoS_2$ doping, resulting in a perfect band matching with $MoS_2$.[21,22] With 80 V gate voltage, the work function of graphene could be shifted up ~4.15 eV,[23] which is lower than the electron affinity of $MoS_2$ (~4.2 eV).[24] The tuneable work function provide excellent band match, leading to a barrier-free contact. Second, compared with conventional metal-$MoS_2$ system, our dry-transfer integration strategy does not involve E-beam lithography and metal deposition process on top of $MoS_2$ contact area, which could damage the underlying $MoS_2$. Unlike metals, graphene is highly inert and stable without any diffusion or reaction with $MoS_2$.[18,25] This non-damage *van der Waals* bonding provide an atomically sharp and ultraclean interface between graphene and $MoS_2$ to minimize defects and interface charge trapping states and prevent Fermi level pinning, which dominates the contact behaviour in conventional metal-MoS$_2$ system.[16,18] It should be noted that graphene has previously been used as contact electrodes on top of MoS$_2$[26-29], but typically with obvious barrier likely due to lithography induced residue at the interface and partial screening of electrical field by MoS$_2$ (especially thick layers)[30].

We now examine the transfer characteristic of these devices under different temperature. Figure 3a shows a temperature dependent transfer curve of monolayer device and the corresponding temperature dependent sheet conductivity ($\sigma$) is plotted in Figure 3b, providing useful information about whether a sample is metallic or insulating. Our studies show that an insulating behaviour ($\sigma$ decrease with T) that persist until V$_g$~50V. With larger gate voltage, the conductivity increases with decreasing temperature, indicating monolayer MoS$_2$ enters a metallic state with critical $\sigma$~e$^2$/h. These observations indicate the presence of a metal$-$insulator transition (MIT) in the sample, consistent with previous observations for monolayer MoS$_2$.[12,20] Benefited from the barrier-free contact, we have for the first time observed such MIT in in a 2-terminal monolayer MoS$_2$ FETs, while can only be seen previously in four terminals measurement to exclude large contact contribution at low T.  In multi-layer MoS$_2$ devices, similar MIT behaviour is also observed (Fig. 3c,d). The difference is that the critical gate voltage decrease to ~20 V. This may be originated from larger carrier concentration of multi-layer MoS$_2$, as well as the smaller band gap (1.2 eV compared with 1.8 eV for monolayer), making devices easier to enter metallic state.

With the barrier-free contact, it is possible to approach the optimized device performance limited by the intrinsic carrier mobility. Here the two-terminal extrinsic FET mobility can be extracted from transconductance using equations $\mu$=[dI$_{ds}$/dV$_{bg}$] $\times$ [L/(WC$_i$V$_{ds}$)], where L/W is the ratio between channel length and width (~between 0.5 to 2 in all devices in Figure 4) and $C_i$ = $1.15 \times 10^{-8}$ F cm$^{-2}$ is the capacitance between the channel and the back gate per unit area (300 nm thick $SiO_2$). To gain further insight into the charge scattering mechanism, we have log-plotted the field-effect mobility of monolayer and multi-layer $MoS_2$ device as a function of temperature. In both monolayer device (Fig. 4a) and multilayer $MoS_2$ devices (Fig. 4d), the mobility increases with decreasing temperature. In the phonon limited temperature range (100-300K), the mobility can be fitted to the expression $\mu \sim T^{-\gamma}$, with the exponent $\gamma$ around 0.66 to 0.84 for monolayer device and 1.26 to 1.74 for multi-layer device. A power law dependence with a positive exponent is indicative of a phonon scattering mechanism, which is consistent with other materials that show band-like transport such as graphene, $Bi_2Te_3$ and other layered materials.[31-33] Further decreasing T from 100 K to 50 K, the mobility of multilayer devices drops due to impurity scattering, consistent with previous calculation and measurements.[12,19,34]

To further optimize device performance, minimize the extrinsic environmental scattering effect and probe the mobility-limiting scattering mechanism, we have used borne nitride (BN) as the encapsulation layer. BN is integrated into the system by picking-up dry transfer technique and edge contact method,[35] with no polymer residue involved between the *van der Waals* stacking. Figure 4b, e shows the mobility versus temperature log-plot of the monolayer and multilayer $MoS_2$ devices encapsulated by a top-BN (around 30 to 50 layers), respectively. Compared with Figure 4d, the multi-layer device shows similar trend with an increase of $\gamma$ (1.92 to 2.12) (Fig. 4e), approaching the value for bulk $MoS_2$.[36] On the other hand, the mobility of monolayer device keep increasing with decreasing temperature with no apparent saturation (Fig. 4b), suggesting that the impurity scattering in monolayer can be suppressed using top-BN encapsulation and the phonon scattering dominates at all temperatures in this case. Overall, compared to the devices with no BN encapsulation, the mobility of the top-encapsulate devices increase ~110% for monolayer and ~30% for multi-layers. The much smaller improvement in multilayer devices can be understood by the existence of top $MoS_2$ layer, already functioning as a screening layers for extrinsic absorbents, resulting less effect for top-BN integration. Furthermore, we have used both bottom and top BN to encapsulate devices in a sandwich structure. Monolayer $MoS_2$ shows similar trend with further mobility increase up to 328 cm$^2$/Vs (Fig. 4c), while the multiple $MoS_2$ demonstrate different curve trend (Fig. 4f) compared with Figure 4e. In Figure 4f, the mobility of multilayer device keep increasing with decreasing temperature without saturation or previous peak around 80 K, indicating the quenching of homopolar mode[12,36] and impurity by bottom BN encapsulation. This suggest that the bottom substrate phonon scattering could be a importation limitation for thick $MoS_2$ device and the mobility can thus increase up to 650 cm$^2$/V s by reducing this effect (Fig. 4f). The quenching of surface-phonon scattering can also be confirmed by the decreased γ value compared with Figure 4e, which is a strong indicative of optical phonon damping. Furthermore, by comparing the monolayer devices (Fig. 4a-c) with multilayer devices (Fig. 4d-f), the exponent γ of monolayer measured here overall is considerable lower than the multilayer, suggesting the difference of electron-phonon coupling in monolayer due to the shift of band valley from Γ–K to K, K'.[36] We would like to note this BN/$MoS_2$/ BN sandwich structure could only be achieved by using graphene as electrode due to its atomic thickness, further highlighting the importance of the graphene contact electrodes architecture.

Although with minimized contact barrier, we note that the contact resistance (including graphene resistance) could still limit the apparent carrier mobility determined from two-terminal transistor measurement. To further reduce the impact of contact resistance and approach the intrinsic field effect mobility limit, we have fabricated a long channel device with large L/W ratio (35/1.7 µm) using BN/MoS$_2$/graphene/BN sandwiched structure (Fig. 5a). The MoS$_2$ here have a thickness of 5 layers. Figure 5b, c shows the optical image of a long MoS$_2$ strip after picked up by large BN and final device after fabrication, respectively. The transfer characteristic with different temperature is plotted in Figure 5d. We can observe that the ON-current increase very rapidly with the decreasing temperature, where the I$_{on}$ at 1.9 K is 12 times larger than I$_{on}$ at 300 K. Compared with regular device geometry (W/L~1), the much larger temperature modulation I$_{1.9\ K}$/I$_{300\ K}$ here strongly indicate less contribution of contact resistance in this long channel device. Further contact engineering are needed to reduce the contact resistance as well as keep its Ohmic barrier free behaviour. The field effect mobility can be extracted from Figure 5d using the same equation mentioned before. **Importantly, the temperature dependent mobility of this device shows that the two-terminal extrinsic mobility can reach up to 1300 cm$^2$/Vs at 1.9 K (Fig. 5e), which represents a record high extrinsic field effect mobility determined from two-terminal measurement, to the best of our knowledge.**

In summary, we have demonstrated that, by using graphene contact electrode with atomically clean interface, a barrier-free contact to MoS$_2$ can be achieved with Ohmic linear I-V output behaviour in all temperature down 1.9 K, which has not been achieved before using other contact methods. Benefited from transparent Ohmic contact, we observed two terminals metal-insulator-transition (MIT) in both monolayer and thick devices, demonstrate a record high extrinsic (2-terminal) mobility over 1300 cm$^2$/Vs. We believe our strategy of using graphene as barrier free contact could shed light on contact engineering in atomically thin 2D transistors and other conventional transistors in general. It could open up a new pathway toward future electronics and low temperature quantum transport in MoS$_2$ and other 2D materials.

**Methods**

**Device fabrication.** The graphene strip (as electrodes) were first peeled on the polymer stack PMMA/PPC (polypropylene carbonate) spun on a silicon wafer, and then transfer it to the boron nitride (as bottom BN) flake sitting on the 300 nm oxidized silicon substrate, following by slowly peeled off the polymer from BN under room temperature. Since the van der Waals force between graphene and BN is stronger than the graphene and polymer stack, graphene will stay on BN while slowly peeling off the polymer. Subsequently, another BN flake (as Top BN) was peeled on PMMA/PPC stack, this top BN was used to pick up $MoS_2$ from another substrate. As the MoS2 was picked up, the entire polymer stack (with Top BN/$MoS_2$) are aligned and transferred onto the as-made graphene/bottom-BN substrate. No solvent was involved during the entire transfer process to ensure ultraclean and atomic sharp contact between graphene and $MoS_2$.

**Microscopic and electrical characterization.** The cross-sectional TEM sample was prepared by focused ion beam cutting and was characterized by a TF20 TEM operating at 300 kV. The D.C. electrical transport studies were conducted with a probe station at room temperature in vacuum with a computer-controlled analogue-to-digital converter. Temperature dependent measurement were conducted using (physical properties measurement system) PPMS by Quantum Design with lowest temperature to 1.9 K.

**Figures and Legends**

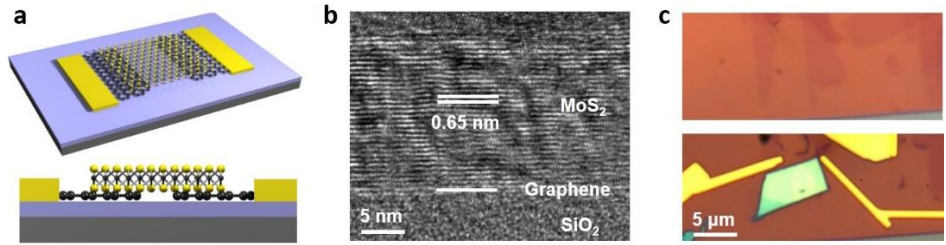

**Figure 1| Device schematics and characterization. a**, Perspective and cross-sectional view of MoS₂ device structure with graphene electrodes. **b,** TEM cross-section image of the graphene MoS₂ interface, indicating the ultraclean and sharp interface of our dry transfer technique. **c,** Optical image of the two graphene stripes peeled close to each other (top panel) and the final device after MoS₂ transfer and metal electrode (bottom panel).

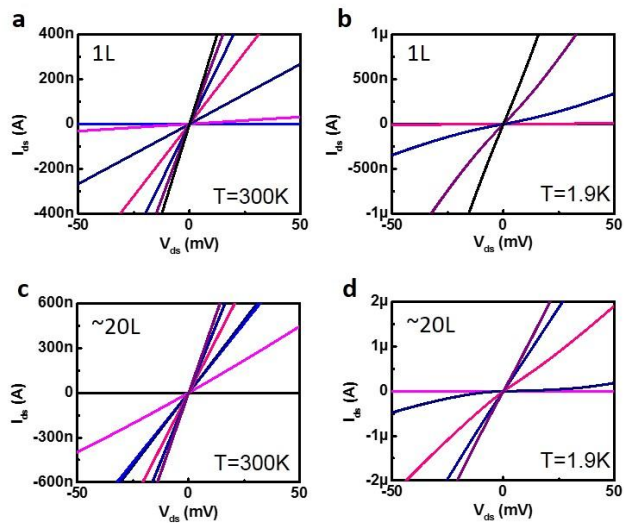

**Figure 2| Output characteristic behaviour of monolayer device and multi-layer MoS₂ device contacted by graphene. a**, **b**, Output characteristic of monolayer MoS₂ device at room temperature (**a**) and low temperature 1.9 K (**b**). Linear I-V behaviour is observed in both cases. Gate voltage is from -60 V to 80 V with 20V step. **c, d,** Output characteristic of multilayer MoS₂ device at room temperature (**c**) and low temperature 1.9 K (**d**). Linear I-V behaviour is observed in both cases. Gate voltage is from -60 V to 80 V with 20V step.

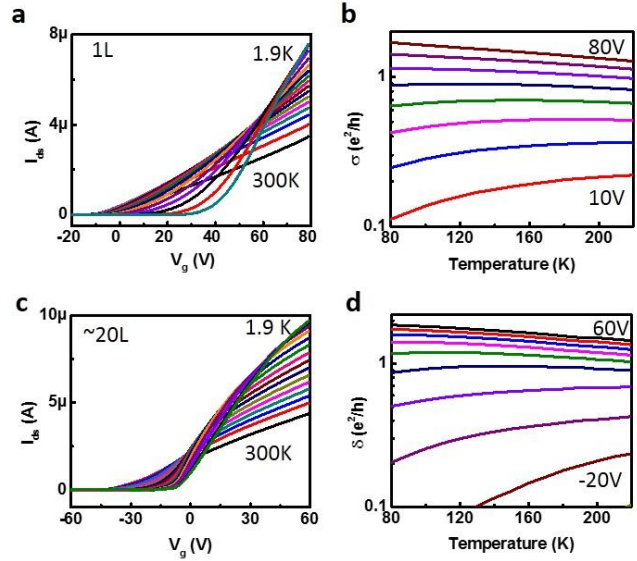

**Figure 3| Transfer behaviour of monolayer device and multi-layer MoS₂ device contacted by graphene. a,** Transfer characteristic of monolayer device**,** at various temperatures (300 K to 1.9 K, 20 K step). $V_{ds}$ is 100 mV. **b,** Transfer characteristic of multilayer device**,** at various temperatures (300 K to 1.9 K, 20 K step). $V_{ds}$ is 100 mV. **c,** The sheet conductivity of monolayer device shown in **a,** with various temperature (-20 V to 60 V, 10V step). **d,** The sheet conductivity of multilayer device shown in **c,** with various temperature (-20 V to 60 V, 10V step).

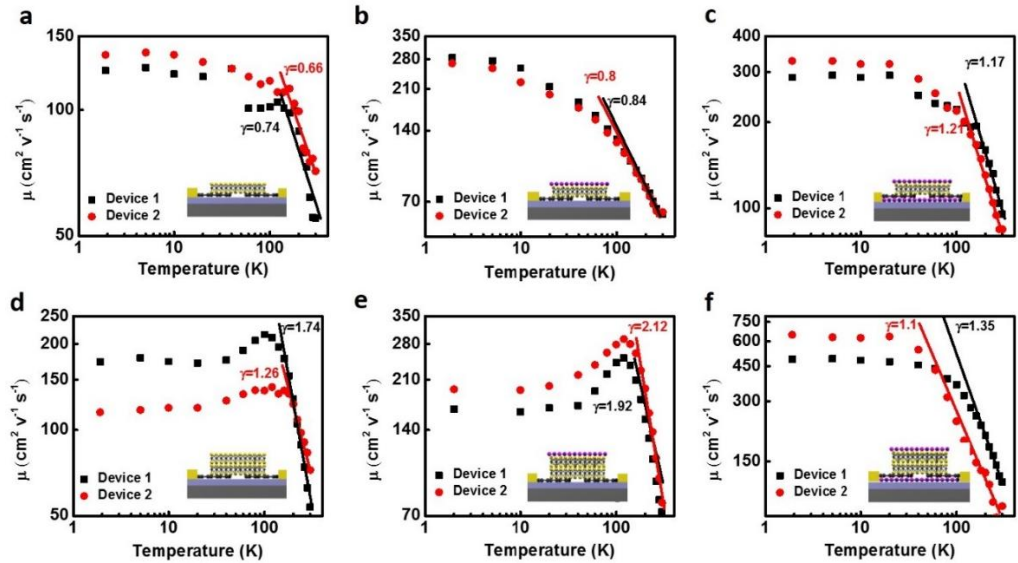

**Figure 4| Mobility engineering by BN encapsulation. a**, **d**, Extrinsic field effect mobility for monolayer and multilayer MoS₂ devices under various temperature (300 K to 1.9 K), linear fit is used in phonon control region (100 K to 300 K) to extract γ. **b**, **e**, Extrinsic field effect mobility for monolayer and multilayer devices under various temperature with top BN encapsulation. **c**, **f**,

Extrinsic field effect mobility for monolayer and multilayer $MoS_2$ devices under various temperature with bottom and top BN encapsulation, forming a sandwich structure.

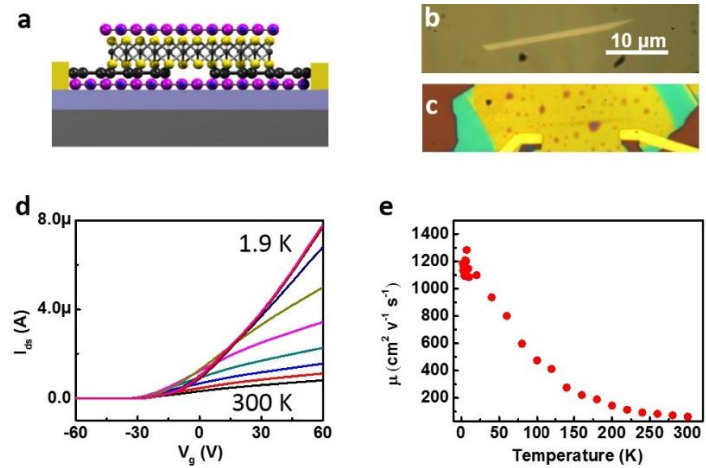

**Figure 5| Long channel sandwiched device to reduce the effect of contact resistance. a**, Schematics of BN/graphene/$MoS_2$/ BN sandwiched structure with edge graphene contact. **b**, Dark field optical image of $MoS_2$ stripe picked up by BN layer. **c**, Device image after device fabrication with edge contact. **d**, Tranfer characteristic of device shown in **c**, large $I_{1.9K}/I_{300\ K}$ can be observed in this device, indicating the large contribution of contact resistance in regular device geometry. **e**, Extrinsic field effect mobility of the device shown in **c** as a function as temperature, with the highest mobility over 1300 cm$^2$/Vs.